\documentstyle[twocolumn,psfig,aps]{revtex}
\begin{document}
\draft
\twocolumn[\hsize\textwidth\columnwidth\hsize\csname @twocolumnfalse\endcsname
%
%
% Title Page
%

\title{ Stripes in the Ising Limit \\
 of Models for the Cuprates }

\author{G. B. Martins$^1$, C. Gazza$^2$, and E. Dagotto$^1$}

\address{$^1$ National High Magnetic Field Lab and Department of Physics,
Florida State University, Tallahassee, FL 32306, USA}

\address{$^2$ Instituto de F\'{\i}sica Rosario (CONICET) and Universidad
Nacional de Rosario, 2000 Rosario, Argentina}

\date{\today}
\maketitle

\begin{abstract}
The hole-doped standard and extended t-J models on ladders with anisotropic
Heisenberg interactions are studied computationally 
in the interval $0.0 \leq \lambda \leq 1.0$ ($\lambda=0$, Ising; $\lambda=1$,
Heisenberg). It is shown that the approximately half-doped 
stripes recently discussed at $\lambda=1$ 
survive in the anisotropic case ($\lambda$$<$1.0), particularly in the ``extended'' model. 
Due to the absence of spin fluctuations in the 
Ising limit and working in the rung basis, a simple picture 
emerges in which the stripe structure can be mostly constructed from 
the solution of the t-J model on chains. 
A comparison of results in the range $0.0 \leq \lambda \leq 1.0$ 
suggests that this picture is valid up to the Heisenberg limit. 

\end{abstract}
\vskip2pc]
\narrowtext

\medskip

In recent years, evidence is accumulating that at least in one
family of high-temperature superconducting
compounds ($\rm La_{2-x} Sr_x Cu O_4$) one-dimensional (1D)
charged stripes are formed upon hole doping of the 
insulating parent material\cite{tranquada}. Although the presence of stripes
in other compounds such as $\rm Y Ba_2 Cu_3 O_{6+\delta}$
is still controversial\cite{dai}, a large theoretical effort has focussed
on the search for stripes in models for the cuprates.
Early results reported stripes 
in Hartree-Fock treatments of the Hubbard model and also in the
phase-separated regime of the t-J model upon the introduction of
long-range Coulomb interactions\cite{zaanen,emery}.  
However, these stripes have
a hole density $\rm n_h$=1.0 in contradiction with the $\rm n_h$=0.5
density found experimentally\cite{tranquada,uchida}. 
Improving upon this situation, recent studies of the t-J model\cite{white,moreo}
reported stripes with $\rm n_h$$<$1.0 at intermediate values of
J/t, sometimes described as a condensation of d-wave
pairs\cite{white2}. However, the
mechanism leading to $\rm n_h$$<$1.0 stripes, and
even the presence of a striped ground-state in the t-J
model at intermediate couplings, is still under
discussion\cite{contro}. 
In addition, evidence is accumulating that the pure t-J model
is not sufficient for the cuprates, and its ``extended'' version with
hopping beyond nearest-neighbor sites\cite{eder,martins} is needed to explain
PES results for the insulators\cite{arpes}.

Very recently, indications of half-doped stripes have been found 
by Martins et al. in the
extended t-J model\cite{prl}. They were also observed at small
J/t in the standard t-J model, in both cases in regimes where
two holes do not form bound states. This led to a novel rationalization
of stripes as the natural way in which spin-charge separation is
achieved in two-dimensional systems\cite{prl}, 
similarly as in the phenomenological
``holons in a row'' picture\cite{zaanen}. 
Moreover, stripes appear to emerge
directly from the one-hole properties of the insulator\cite{prl}, 
where holes have strong ``across the hole'' 
antiferromagnetic (AF) correlations\cite{martins,white3} in their 
(frustrated) effort to achieve spin-charge separation, similarly 
as it occurs in 1D systems\cite{shiba}. 

In spite of this progress, more work
is needed to understand these complex striped states. An
important issue is the role played by fluctuations in the spin
sector. While the presence of a spin tendency to form an AF 
background is crucial for stripe formation, it is unknown whether 
the fine details of the spin sector (such as the presence of
low-energy excitations) are important for its
stabilization. 
%Can stripes be formed in a spin-gapped background or
%Are low-energy spin fluctuations important for a striped ground-state?
To address this question, here a computational study is reported where
the spin interaction contains an Ising anisotropy. Our main result is that
stripes survive the introduction of this anisotropy, and a fully SU(2)
symmetric interaction is not needed for stripe
formation. This result is in agreement with recent
retraceable-path calculations for the t-J$_z$ model\cite{castroneto}.

The anisotropic extended t-J model is defined as
%\begin{eqnarray}
%\rm
$$
\rm H = J \sum_{\langle {\bf ij} \rangle} 
[{{S^z}_{\bf i}}{{S^z}_{\bf j}}
+ {{\lambda}\over{2}}({{S^+}_{\bf i}}{{S^-}_{\bf j}}
+{{S^-}_{\bf i}}{{S^+}_{\bf j}}) -{{1}\over{4}}n_{\bf i}n_{\bf j} ]
$$
$$
\rm  - \sum_{ {\bf im} } t_{\bf im} (c^\dagger_{\bf i} c_{\bf m} +
h.c.),~~~~~~~~~~~~~~~~~~~~
\eqno(1)
$$
%\end{eqnarray}
where $\rm t_{\bf im}$ is t(=1) for nearest-neighbors (NN) hopping
between sites ${\bf i}$ and ${\bf m}$, t$'$ for 
next NN, t$''$ for next-next NN, and
zero otherwise. The anisotropy in the spin sector 
is controled by 
$\lambda$ ($\lambda=0$, Ising; $\lambda=1$, Heisenberg). 
The rest of the notation is standard. t$'$=-0.35 and
t$''$=0.25 are believed to be 
relevant to explain PES data\cite{eder,martins,arpes}.
The computational work is carried out using the
Density Matrix Renormalization Group (DMRG)\cite{white,error} and
Lanczos\cite{review} techniques, as well as an algorithm using
a small fraction of the ladder rung-basis 
(optimized reduced-basis approximation, or ORBA\cite{orba}).
Results are presented in (i) the
small J/t region with t$'$=t$''$=0.0, and (ii) small
and intermediate J/t with nonzero 
t$'$ and t$''$,  in both cases working at $\lambda$=0.0, 0.25, 
0.50 and 1.00. These two
regions (i) and (ii) have similar physics\cite{martins}, and the extra
hoppings are expected to avoid
phase separation\cite{contro,tohyama}.

To start our investigation, let us analyze  
the one-hole properties of model Eq.(1).
Previous studies found a robust AF correlation
between spins 
 across-the-hole (C$_{AH}$)\cite{martins,white3}, namely between
the spins separated by two lattice spacings located on both sides of a
hole (working in the reference frame of the latter).
This curious feature
was interpreted as a short 
distance tendency toward spin-charge separation\cite{martins},
and it is believed to be crucial for stripe formation\cite{prl}.
It is important to clarify if similar correlations 
are still present in the anisotropic case. 
For this purpose, here
 4$\times$4 and 4$\times$6 clusters were used with periodic boundary 
conditions (PBC), as well as cilindrical boundary conditions (CBC),
(open boundary conditions (OBC) along legs and PBC 
along rungs\cite{white}), for several different couplings\cite{comm1}. 
The results for C$_{AH}$ in the one-hole lowest energy state at
several momenta are in Fig.1a using a PBC 4$\times$6 cluster,
at a representative coupling set. At $\lambda$=1.0 all the correlations
are negative, compatible with the results in
Refs.\cite{martins,prl}. As $\lambda$ is reduced, the magnitude of 
C$_{AH}$ decreases but it remains negative for the momenta of most relevance,
namely $(\pi/3,\pi/2)$ (the closest on the 4$\times$6 cluster to
$(\pi/2,\pi/2)$) and $(0,\pi)$-$(\pi,0)$. In Fig.1b the AF spin
correlations are shown at $(0,\pi)$ and $\lambda$=0.0. There is a good
agreement
with the results observed in the $\lambda$=1.0 limit\cite{martins}. 
Similar conclusions are reached in the $(\pi,\pi)$ case (Fig.1c)
at moderate anisotropy, and for $(\pi/3,\pi/2)$ (not shown) 
although in this case 
the correlation along the short direction is approximately zero.
The across-the-hole correlations remain even with Ising
anisotropy.

\begin{figure}
\begin {center}
\mbox{\psfig{figure=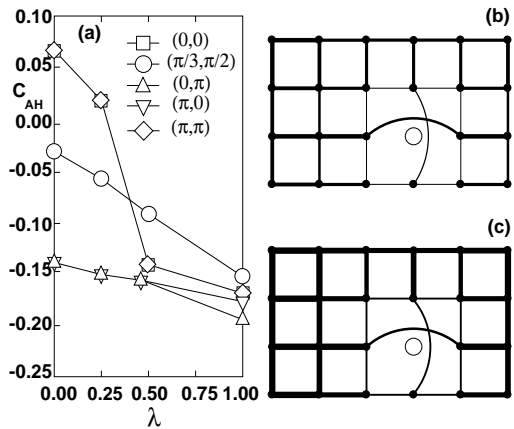,height=2.2in,width=2.64in}}
\end{center}
\caption{(a) Exact Diagonalization
calculations of across-the-hole spin-spin correlations 
(C$_{AH}$) vs. $\lambda$ 
for several momenta on a 4$\times$6 cluster with 1 hole, PBC,
J=0.2, t$'$=-0.35, 
and t$''$=0.25. The results shown are for the 
long direction. (b) AF spin-spin correlations for (a) at ($\pi$,0) 
and $\lambda$=0.0. (c) AF spin-spin correlations for (a) 
at ($\pi$,$\pi$) and $\lambda$=0.5. In (b) and (c) the hole is projected
from the ground-state to the site shown, and the dark lines represent
AF spin-spin correlations with a thickness proportional to its absolute
value.
} 
\label{fig1}
\end{figure}

%
%
%Regarding the dependence of C$_{AH}$ with $\lambda$, the results showed 
%that for most of the different sets of 
%parameters used there was a variation on C$_{AH}$ from AF 
%to ferromagnetic   with decreasing $\lambda$ for $\lambda \leq
%0.5$. This happens mainly for momenta 
%(0,0) and ($\pi,\pi$) (PBC) and k$_y$=0, $\pi$ (CBC). It is also clear 
%in Fig. 1a, where exact diagonalization (ED) results for C$_{AH}$ are 
%shown for a 4$\times$6 cluster with 1 hole, PBC in both directions, 
%for J=0.2, t$'$=-0.35, and t$''$=0.25, that there is an upward 
%trend on C$_{AH}$ with decreasing $\lambda$ for all momenta. 
%
On the other hand, for momenta (0,0)
and $(\pi,\pi)$ with PBC (and also at k$_y$=0 and $\pi$, with CBC)
a transition to a $ferromagnetic$ (FM) correlation occurs at
small $\lambda$. This tendency is dangerous for the stripe formation,
which will not occur in a spin polarized background.
Calculating the NN spin-correlations {\it without} hole projection, i.e.
involving all the links of the cluster for 
several couplings and $0.0 \leq \lambda \leq 1.0$, similar 
trends were observed, including the change of sign for $\lambda 
\rightarrow 0.0$.
This suggests that the C$_{AH}$ $\lambda$-dependence  at (0,0) and
($\pi,\pi$) for the one-hole results, can be 
ascribed to an overall decrease of the AF tendency on the clusters
studied, which is replaced by a tendency toward ferromagnetism. 
%
%Even though an AF C$_{AH}$ 
%could still be dynamically generated by the hopping terms, 
%a cluster with no overall tendency to AF order in the one-hole sector 
%will likely tend to suppress the across-the-hole structure\cite{barnes}. 
%
A comparison of data at several couplings and lattice sizes
%
%for the same couplings on 
%different clusters (4$\times$4 and 4$\times$6), 
%and on the same cluster (4$\times$4) at different couplings 
%(J=0.2, t$'$=-0.35, t$''$=0.25; and J=0.4, same t$'$ and t$''$) 
%gives 
supports this interpretation. Fortunately, the
results are also in agreement
with the early analysis by Barnes et al.\cite{barnes} that reported
a reduction in the FM tendency upon the increase of the
cluster size for the Ising limit of the t-J model. Thus, it is believed
that the FM state found in some sectors of the one-hole problem
at finite J will disappear as the clusters grow\cite{comm2}.
In fact, if a compromise between couplings and 
cluster size is followed to avoid the FM region, the qualitative ``shape''
of the 
$\lambda$=1.0 one-hole wave-function  
can be preserved as $\lambda$$\rightarrow$0.0. 
%For example, 
%C$_{AH}$ is $\approx$0.0 on a 
%4$\times$4 cluster at $\lambda$=0.0 and momentum (0,$\pi$) using 
%the couplings of Fig.1, but
%this value climbs to $\approx$$-0.15$ 
%on a 4$\times$6 cluster (see Fig. 1a). 
In conclusion, the FM tendency 
in some subspaces of the one-hole sector is expected not to be detrimental to
stripe formation, and studies 
with more holes shown below support this view. Nevertheless, care must
be taken with the stripe-FM competition in these systems.

% less hole mobility 
%and mainly bigger clusters tend to generate a more robust C$_{AH}$ 
%for $\lambda \approx 0.0$.  
%
%As discussed in previous papers for $\lambda$ = 1.0 \cite{prl} 
%the accros the hole AF correlations are a 
%robust feature in a region of parameter space that is compatible 
%with PES results for undoped 
%cuprates. As $\lambda$ is varied, specially for $\lambda \simeq 0.0$, 
%a comparison of results 
%for different clusters and couplings shows that, 
%
%And the jump of C$_{AH}$ 
%from AF to FM for (0,0) and ($\pi$,$\pi$), 
%seen in Fig 1a, has a bigger magnitude in a 4$\times$4 cluster 
%at the same coupling.

\vspace{-0.12in}
\begin{figure}
\begin {center}
\mbox{\vspace{-0.25in}\psfig{figure=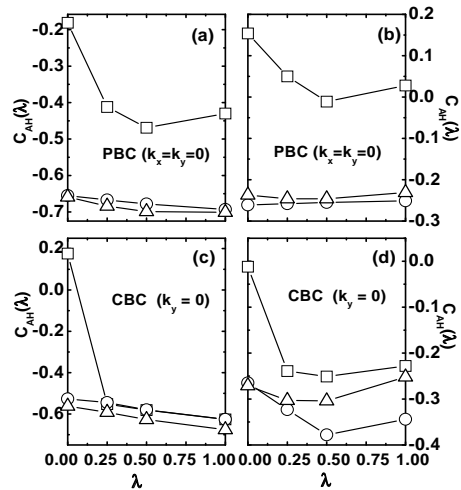,height=3.0in,width=2.36in}}
\end{center}
\vspace{-0.25in}
\caption{Spin-spin correlations vs $\lambda$ for the most probable 
ground-state hole 
configuration (stripe) using 4$\times$4 clusters with 2 holes, PBC and 
CBC. (a) Spin-spin correlations across-the-hole calculated along the
stripe; PBC were used in both directions. 
Squares are for J=0.2 and t$'$=t$''$=0.0; circles for J=0.2, t$'$=-0.35,
and t$''$=0.25; and 
triangles for J=0.4, t$'$=-0.35, and t$''$=0.25. In all three cases
k$_x$=k$_y$=0. (b) Same as 
(a) but now the spin-spin correlations across-the-hole are calculated
across the stripe. (c) Same as 
(a) but now using CBC k$_y$=0 (the stripe runs along the PBC direction).
%Note that for 
%the standard t-J model (squares) FM correlations develop in 
%the Ising limit, as would be expected for J=0.2 
%on a 4$\times$4 cluster (Fig.1).
(d) Same as (b) but 
now using CBC k$_y$=0.
} 
\label{fig2}
\end{figure}

Consider now two holes on 4-leg ladders. Similarly as 
in Ref.\cite{prl}, stripes are formed with the two
holes mainly located at two lattice spacings along the rung. Then,
correlations along and across the stripe must be considered separately.
As in Fig.2, 
the introduction of a second hole on a 4$\times$4 cluster using the
$extended$ t-J model
stabilizes robust C$_{AH}$ both along and across the stripe, in PBC and 
CBC, even for values of $\lambda$ where the one-hole system did not have 
a robust AF C$_{AH}$. For the standard
t-J model at J=0.2, t$'$=t$''$=0.0, also shown in Fig.2, the situation
is less clear and in some cases the correlations become FM
at $\lambda$=0.0, but in most situations it remains AF. 
In the extended t-J model the stripe
tendency is clearly stronger than in its standard version\cite{comm4}.

% can be rationalized the following way: As shown
%above, for one-hole calculations, the absence of robust AF C$_{AH}$ for 
%small $\lambda$ in some clusters, and for some couplings, can be
%associated to a decrease in the overall AF order with decreasing
%$\lambda$. But even for the cases where C$_{AH}$ is FM its value is still 
%considerably smaller than a typical non-across-the-hole next-nearest-neighbor 
%spin-correlation in the same system. This indicates that, even though 
%C$_{AH}$ is positive, the mechanism that generates an AF C$_{AH}$ is 
%still operative, although not completely effective. As this mechanism 
%is basically related to optimization of the kinetic energy of the hole, 
%once another hole is introduced it becomes more effective, since, as 
%shown on ref. \cite{prl}, the 'linking' of C$_{AH}$ from nearby holes 
%avoids spin frustration and once the holes form a stripe the overall AF order 
%of the background can be restored, with the help of an
%antiphase-domain-wall. This way, fully AF accros the hole spin 
%correlations (along and across the stipe) can be stabilized.

\vspace{-0.1in}
\begin{figure}
\begin {center}
\mbox{\psfig{figure=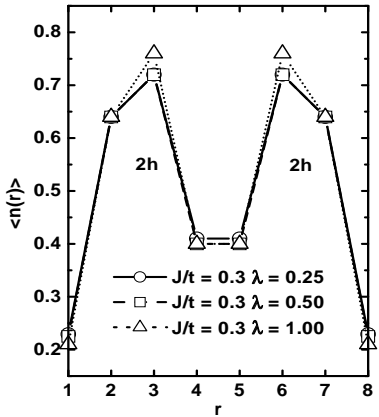,height=2.7in,width=2.3in}}
\end{center}
\vspace{-0.3in}
\caption{Rung hole density $\langle$n(r)$\rangle$ vs rung index r
using DMRG (243 states) on a 4$\times$8 cluster with 4 holes, 
CBC (OBC along the direction shown, with invariance under reflections
assumed). 
%Note that the stripes are not affected by the fluctuations 
%in the spin-background regulated by $\lambda$$\neq$0.0
} 
\label{fig3}
\end{figure}

Previous studies at $\lambda$=1.0 showed the 
coexistence of two or more n$_h$=0.5 stripes 
at hole density x=1/8 on 4-leg ladders
as the number of holes increase in a 
CBC cluster. It is important to show that this 
stripe phase survives the decrease of the spin 
fluctuations. That this is the case can be observed in Fig. 3 
where DMRG results for the rung density
$\langle$n(r)$\rangle$ vs the rung label r are presented
for a 4$\times$8 cluster with 4 holes (J=0.3, t$'$=t$''$=0.0; CBC). 
Figure 3 shows that from
$\lambda$=1.0 down 
to $\lambda$=0.25 the two n$_h$=0.5 stripes are virtually unaltered.
On the other hand, at $\lambda$=0.0 the FM tendency already discussed
in the 2-holes case of Fig.2 prevent the formation of stripes in the
standard t-J model. This problem does not occur in the extended version
with t$'$$<$0 and t$''$$>$0.

Let us now analyze the spin structure around the stripe once
the spin fluctuations are fully turned off ($\lambda$=0.0). 
In Fig.4a, a stripe 
configuration with a clear $\pi$-shift across-the-stripe is shown for 
a CBC 4$\times$6 cluster with 2 holes at $\lambda$=0.0. 
This dominant hole configuration was projected out of the ground state 
of an 
ORBA calculation where $\approx$ 10$^6$ states were kept in the rung basis. 
The couplings (J=0.2, t$'$=0.0, t$''$=0.25) differ slightly 
from those in Fig.2, but the results are representative 
and they are similar to those found 
at $\lambda$=1.0\cite{correl}. 

\begin{figure}
\begin {center}
\mbox{\psfig{figure=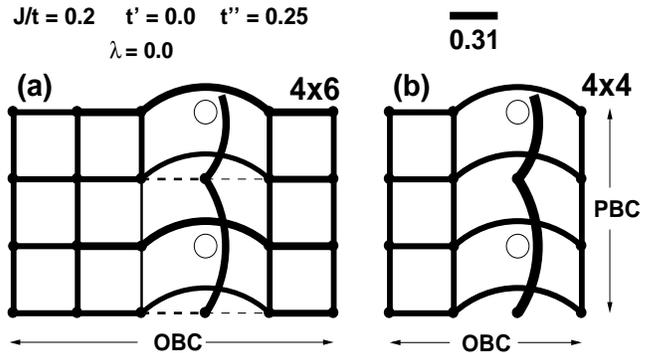,height=1.8in,width=3.3in}}
\end{center}
\caption{
(a) Spin-spin correlations for 2 mobile holes projected at their
most probable configuration (circles) in the 2 holes ORBA ground-state
of a 4$\times$6 cluster using CBC (k$_y$=0), J=0.2, t$'$=0.0, t$''$=0.25, 
and $\lambda$=0.0. Full lines
indicate AF correlations (thickness proportional to
absolute value); dashed lines indicate ferromagnetic correlations.
(b) AF spin-spin correlations of the CBC 4$\times$4 cluster with 2 holes 
projected at their most probable distance. This result was obtained 
keeping just the 8 rung-basis 
states with the highest weight in the ground state 
(out of a full Hilbert space with 102,960 states). 
Same couplings and momentum as in (a). This approximation emphasizes 
the 1D character of the stripe; note also that the stripe formed by this
procedure 
is completely decoupled from the rest of the cluster.
} 
\label{fig4}
\end{figure}

It is expected that the 
suppression of spin fluctuations would simplify the description of a ground
state like the one in Fig.4a, either in the rung- or S$z$-basis, 
since less states are needed to represent the spin background. 
Indeed the combination of CBC with the use of 
a rung basis (along the PBC direction) leads to a fairly simple
description of such a state. This can be observed in
Fig.4b, where the spin-correlations 
were calculated on 
a CBC 4$\times$4 cluster with 2 holes (same coupling
as in Fig.4a) using only the $eight$ highest weight rung-basis
states out of the full ground state of the system, that has a total of
102,960 states. 
The results are very similar to those found with the full
ground-state and they capture
the basic physics contained in the full calculation. The picture 
that emerges is the following: the spin-correlations along the stripe are
maximized, i.e., the two spins on it are locked in a singlet, 
implying that their correlations 
with the rest of the spins vanish. This means that in the ``snapshot'' 
of the ground-state (Fig.4b) the stripe 
is disconnected from the rest of the cluster, emphasizing its 1D
character. In fact the only rung-state that contains holes in the eight most
dominant states kept in Fig.4b
is the ground-state of the 2-hole sector of a simple 
4-site ring calculation. This
establishes a strikingly simple connection of the stripe problem 
with a truly 1D calculation. The ``recipe'' to construct a good
representation of the stripe state is to consider the solution of
half-doped 1D chains as a stripe, and antiferromagnetically
couple the rest of the spins of
the plane simply as
if those stripes would be absent (thus generating the $\pi$-shift
across the stripes). This is a natural generalization to two-dimensions (2D)
of the 1D spin-charge separation concept, as emphasized in Ref.\cite{prl},
where the spin portion of the wave function is constructed simply
ignoring the holes\cite{shiba}. In 2D now it is the stripes that 
are ignored by the
spins not belonging to them in their wave function.
It is also important to notice that this picture appears to hold, 
in its main aspects, even at $\lambda$=1.0.

  For completeness, in Fig.5a ORBA results
(with $\approx$ 2$\times$10$^6$ states kept in the rung-basis)
are presented
for $\langle$n(r)$\rangle$ vs r, showing two n$_h$=0.5 stripes with two
holes each. 
%In one cluster the stripes contain 2 holes 
%(4$\times$6, $\lambda$=0.0, solid line) and in the other they have
%3 holes each (6$\times$6, $\lambda$=0.5, dashed line). Both calculations
The calculation was performed with CBC.
%, and the standard t-J model (J=0.3). 
In Fig.5b details of the spin-correlations 
are shown, with
%on a CBC 4$\times$6 cluster with 4 holes (J=0.2, t$'$=0.0, t$''$=0.25), 
%where 
the most probable hole configuration projected out of the 
ORBA ground-state. At 
$\lambda$=0.0 a very good convergence in the ORBA method 
can be achieved: starting with initial states where the holes are uniformly
distributed or phase separated, a fast
convergence leads to the stripe 
results of Fig.5b.

\begin{figure}
\begin {center}
\mbox{\psfig{figure=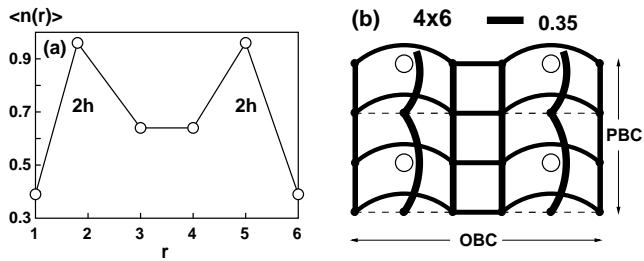,height=1.3in,width=3.3in}}
\end{center}
\caption{
(a) Rung hole density $\langle$n(r)$\rangle$ vs
rung index r 
calculated using ORBA on a 4$\times$6 cluster with 4
holes, J=0.2, t$'$=0.0, t$''$=0.25, and $\lambda$=0.0, using CBC
with k$_y$=0.
% (J=0.3, t$'$=t$''$=0.0,
%and $\lambda$=0.25). 
%CBC in both cases, 
(b) Spin-spin correlations for 4 mobile holes at their
most probable configuration in the 4 holes ORBA ground state
of (a).
%of a 4$\times$6 cluster using CBC (k$_y$=0), J=0.2, t$'$=0.0, t$''$=0.25, 
%and $\lambda$=0.0.
} 
\label{fig5}
\end{figure}

Summarizing, it has been shown that the
striped states recently identified in 
models for the cuprates survive the introduction
of an Ising anisotropy, particularly in the extended t-J model. The important physics that
stabilizes stripes appears to be the competition between  
spins that order antiferromagnetically and
holes that need to modify the spin environment to
improve their movement, leading to an interesting potential extension
into 2D of the familiar 1D spin-charge separation ideas.
The approach discussed here focuses on the small J/t limit, and it appears
unrelated with others based on the large J/t phase separated region.
The fine details of the spin background do not seem important, 
and as a consequence doped stripes should be a general 
phenomenon in correlated electronic systems.

The authors thank NSF (DMR-9814350), the NHMFL (In-house Research Program), 
and MARTECH for partial support.

\vfil
%\vspace{-0.5in}


\begin{references}
\bibitem{comm1} Fortunately, there is a weak 
dependence on boundary conditions: the 
results for CBC calculations, with correlations measured
along the PBC direction, are qualitatively 
the same as for the PBC calculations for all couplings studied. 
The results along the OBC direction, although not very different, show 
some variation when compared to the results along the PBC direction: 
on average C$_{AH}$ is more AF in the OBC direction 
and its dependence with $\lambda$ is less monotonic than in the PBC direction.
\bibitem{comm2} As it occurs in
the Nagaoka state of the one-hole sector of the t-J model, which as the
cluster grows is
stabilized at smaller J/t, reaching the J/t=0.0 limit 
in the bulk\cite{review}.
\bibitem{comm4} 
The results thus far are not in contradiction with 
our recent interpretation of stripe formation as related to the 
one-hole properties\cite{prl}, since this simple argument holds as
long as the overall tendency in the spin sector is toward an
AF state. Once ferromagnetism competes for one hole,
the state that induces stripes upon further doping is now 
an excited state.
\end{references}
\end{document}